\documentclass[11pt]{article}
\usepackage[utf8]{inputenc}
\usepackage[T1]{fontenc}
\pdfoutput = 1

\usepackage{amsmath}
\usepackage{amssymb}
\usepackage[english]{babel}
\usepackage{bbm}
\usepackage{bbold}
\usepackage{braket}
\usepackage{caption}
\usepackage{xcolor}
    \definecolor{darkgreen}{rgb}{0,0.5,0}
    \definecolor{darkred}{rgb}{0.5,0,0}
    \definecolor{darkblue}{rgb}{0,0,0.6}
    \definecolor{purple}{rgb}{0.4,.2,0.7}
\usepackage{comment}
\usepackage{csquotes}
\usepackage[mathcal]{eucal}
\usepackage{graphicx}
\usepackage[hyperfootnotes = true, colorlinks = true, linkcolor = darkblue, citecolor = purple]{hyperref}
\usepackage{mathabx}
\usepackage{mathtools}
\usepackage{pdfsync}
\usepackage{slashed}
\usepackage{subcaption}
\usepackage{tikz}
\usepackage{upgreek}
\usepackage{url}
\usepackage[normalem]{ulem}

\usepackage[margin = 2.2cm]{geometry}
\usepackage[ragged]{footmisc}
\renewcommand{\d}{\mathrm{d}}
\renewcommand{\i}{\mathrm{i}}
\DeclareMathOperator{\Tr}{Tr}
\DeclareMathOperator{\tr}{tr}
\newcommand\stac[2]{\genfrac{}{}{0pt}{}{#1}{#2}}

\begin{document}

\thispagestyle{empty}
\begin{center}
    ~\vspace{5mm}
    
    {\Large \bf 

        Giant graviton expansion from eigenvalue instantons
    
    }
    
    \vspace{0.4in}
    
    {\bf Yiming Chen,$^1$ Raghu Mahajan,$^1$ and Haifeng Tang$^{1}$}

    \vspace{0.4in}

    $^1$ Department of Physics, Stanford University, Stanford, CA 94305-4060, USA 
    \vspace{0.1in}
    
    {\tt ymchen.phys@gmail.com, raghumahajan@stanford.edu, hftang@stanford.edu}
\end{center}

\vspace{0.4in}

\begin{abstract}
Recently, S. Murthy has proposed a convergent expansion of free partition functions and superconformal indices of finite-$N$ purely adjoint gauge theories based on a Fredholm determinant expansion.
This expansion has been dubbed the giant graviton expansion and takes the form of an infinite series of corrections to the $N=\infty$ result, with the $m^\text{th}$ correction being of order $e^{-mN}$.
We show that this expansion can be reproduced using eigenvalue instantons in unitary matrix integrals.
This perspective allows us to get the giant graviton expansion proposed by S. Murthy without the intermediate step of the Hubbard Stratonovich transformation.
\end{abstract}

\pagebreak

A puzzling feature of holographic duality is the overcompleteness of the semiclassical bulk Hilbert space---states which appear orthogonal from the  perspective of the semiclassical bulk theory can have exponentially small overlaps, and complicated superpositions of them can result in null states. 
Trace relations \cite{Balasubramanian:2001nh, Lee:2023iil} and other nonperturbative effects (for instance, due to spacetime wormholes \cite{Marolf:2020xie, Penington:2019kki}) lead to fewer states in the finite $N$ theory compared to the infinite $N$ theory. 
Such effects have been proposed to play important roles in the black hole information problem \cite{Akers:2022qdl}, the error correcting features of holography \cite{Faulkner:2022ada} and the fortuity of supersymmetric black holes \cite{Chang:2024zqi}. 
In this note, we study a non-perturbative effect of this kind, namely the existence of trace relations in finite-$N$ gauge theories with adjoint fields, using a collective field description that is analogous to semiclassical gravity. 
We aim to elaborate how the instanton effects in such models reflect the existence of null states. 

Concretely, we will study superconformal indices of supersymmetric gauge theories, and partition functions of free gauge theories, which are given by unitary matrix integrals, the unitary matrix being the holonomy of the gauge field on the temporal $S^1$ \cite{Sundborg:1999ue, Aharony:2003sx} .
Restricting to $U(N)$ gauge theories with purely adjoint fields, the unitary matrix integrals take the form 
\begin{align}
    Z_N(\mathbf{g}) := \int \d U \, \exp \left( 
    \sum_{k=1}^\infty \frac{g_k}{k} \, \Tr U^k \Tr U^{-k}
    \right)\,,
    \label{zndef}
\end{align}
where $g_k$ are some coefficients that are determined by the temperature and chemical potentials present in the specific problem being studied.
The Haar measure $\d U$ is normalized so that the volume of $U(N)$ group manifold equals one.

We will be interested in studying this integral at finite $N$, and not just at infinite $N$.
The integral (\ref{zndef}) exhibits non-perturbative effects due to trace relations, which have been dubbed ``giant graviton'' effects, following the role played by giant gravitons in the 1/2-BPS sector of $\mathcal{N}=4$ SYM theory \cite{McGreevy:2000cw}.
In \cite{Murthy:2022ien}, a formula was proposed by S. Murthy for general integrals of this type where the finite-$N$ result is written as a series of exponentially small corrections to the infinite-$N$ result.\footnote{
For other results of this nature, proposing similar but different expansions, see e.g. \cite{Gaiotto:2021xce, Lee:2022vig, Lee:2023iil, Imamura:2021ytr, Imamura:2022aua, Arai:2019wgv, Arai:2019xmp, Arai:2020qaj,Beccaria:2023zjw,Beccaria:2024oif,Imamura:2024lkw,Beccaria:2024lbt}.
For implications of this formula for bulk physics in AdS, see \cite{Eleftheriou:2023jxr, Beccaria:2024vfx, Beccaria:2023hip, Choi:2022ovw, Kim:2024ucf}. 
We will only be studying non-perturbative results in the ``confining'' phase of such integrals. For eigenvalue instantons in the ``deconfined'' phase and their bulk interpretation, see for example \cite{Aharony:2021zkr}.
}
Let us state the result.
First, one performs a Hubbard-Stratonovich transformation to define a new integral $\widetilde{Z}_N(\mathbf{t^+},\mathbf{t^-})$ as follows
\begin{align}
    Z_N(\mathbf{g}) &= \int \prod_{k=1}^\infty \frac{\d t_k^+\d t_k^-}{2\pi k g_k} \exp \left( 
    - \frac{t_k^+ t_k^-}{k g_k}
    \right) \widetilde{Z}_N(\mathbf{t^+},\mathbf{t^-})\,, \quad \text{where} \label{hstransformation} \\
    \widetilde{Z}_N(\mathbf{t^+},\mathbf{t^-}) &:= 
    \int \d U \, \exp \left( 
    \sum_{k=1}^\infty \left( 
    \frac{t_k^+}{k} \Tr U^k  + \frac{t_k^-}{k} \Tr U^{-k} 
    \right)
    \right) \, . \label{zntildedef}
\end{align}
Ref. \cite{Murthy:2022ien} then invokes the result of \cite{Geronimo:1979iy,borodin2000fredholm}, which states that
\begin{align}
    \widetilde{Z}_N(\mathbf{t^+},\mathbf{t^-})
    &=  \widetilde{Z}_\infty(\mathbf{t^+},\mathbf{t^-})\left( 1 + \sum_{m=1}^{\infty} \widetilde{G}^{(m)}_N(\mathbf{t^+},\mathbf{t^-})  \right)\,,\quad \text{with } \label{gntildeexpansion} \\
    \widetilde{Z}_\infty(\mathbf{t^+},\mathbf{t^-}) &= \exp \left( 
    \sum_{k=1}^\infty \frac{t_k^+ t_k^-}{k} \right)\,, \quad \text{and} \label{zinftyt} \\
    \widetilde{G}^{(m)}_N(\mathbf{t^+},\mathbf{t^-}) &= (-1)^m \sum_{\stac{\{ r_i \} : N <r_1 < \cdots r_m }{r_i \in \mathbb{Z} + 1/2}}
    \det \begin{pmatrix} 
    \widetilde{K}(r_1,r_1) & \widetilde{K}(r_1,r_2) & \dots & \widetilde{K}(r_1,r_m) \\
    \vdots & \vdots & \ddots & \vdots \\
    \widetilde{K}(r_m,r_1) & \widetilde{K}(r_m,r_2) & \cdots & \widetilde{K}(r_m,r_m)
    \end{pmatrix}  
    \,,  \label{gnmtildedef}
\end{align}
and where $\widetilde{K}$ is an infinite dimensional matrix defined by the generating function
\begin{align}
    \sum_{r,s,\in \mathbb{Z} + \frac{1}{2}} \widetilde{K}(r,s) \, v^r u^{-s} :=
    \frac{\sqrt{v u}}{v-u} \exp \left(
    \sum_{k=1}^\infty 
    \left( 
    \frac{t_k^+}{k} (v^k - u^k) - \frac{t_k^-}{k} (v^{-k} - u^{-k})
    \right)
    \right)\,,\quad  \vert u \vert < \vert v \vert \, .
    \label{ktildegenfn}
\end{align}
Note that all the matrix indices of $\widetilde{K}$ appearing in (\ref{gnmtildedef}) are larger than $N$.
Substituting the expansion (\ref{gntildeexpansion}) into (\ref{hstransformation}), we get the required expansion, which is written as
\begin{align}
    Z_N(\mathbf{g}) = Z_\infty(\mathbf{g}) \left( 1 +   \sum_{m=1}^\infty G^{(m)}_N(\mathbf{g}) \right)\, .
    \label{gnexpansion}
\end{align}

Our goal in this note is to reproduce the expansion (\ref{gnexpansion}) from the point of view of eigenvalue instantons in the matrix integral (\ref{zndef}).

For this we will need to review two more results.
The first is from \cite{Liu:2022olj, Eniceicu:2023cxn} (see, for example, Eq. (5.2) of \cite{Eniceicu:2023cxn}), where it was shown that
\begin{align}
    G^{(m)}_N &= \frac{(-1)^m}{(m!)^2}\int \prod_{i=1}^m\frac{\d u_i}{2\pi \i u_i^{-N}} \frac{\d v_i}{2\pi \i v_i^N}
    \frac{\prod_{1 \leq i < j \leq m} (u_i - u_j)^2(v_i - v_j)^2}{\prod_{i=1}^m \prod_{j=1}^m (u_i - v_j)^2}  \times \nonumber \\
    &\qquad  \times \exp \left( 
    - \sum_{k=1}^\infty \frac{1}{k} \frac{g_k}{1-g_k} \sum_{i=1}^m \sum_{j=1}^m
    (u_i^k - v_i^k)
    (u_i^{-k} - v_i^{-k})
    \right)
    \, .
    \label{danresult}
\end{align}
Our strategy is to show that the integral on the right hand side of (\ref{danresult}) is precisely what one gets by studying the appropriate eigenvalue instanton problem.

\paragraph{Review of eigenvalue instantons in unitary matrix integrals with single-trace potentials.}

The second set of results that we need are about the eigenvalue density and the one-eigenvalue effective potential in the integral (\ref{zntildedef}). 
These results are well known, but we refer the reader to appendix C of the recent paper \cite{Eniceicu:2023cxn} for a recent pedagogical exposition. 
We will be working in the large-$N$ limit, and $1/N$ will only be used as a perturbation parameter appearing in formal series expansions.

\begin{figure}[t!]
  \centering
  \begin{subfigure}[b]{0.4\textwidth}
    \centering
    \begin{tikzpicture}
      \draw (0,0) circle (1.5cm); 
      
      \newcommand{\NumDots}{17} 
      
      \foreach \x in {1,2,...,\NumDots} {
        \pgfmathsetmacro{\Angle}{90 - 360/\NumDots*(\x-1)}
        \fill (\Angle:1.5cm) circle (1.5pt);
        \ifnum\x=1
          \node at (\Angle:1.8cm) {$z_1$};
        \else
          \ifnum\x=2
            \node at (\Angle:1.8cm) {$z_2$};
          \else
            \ifnum\x=3
              \node at (\Angle:1.8cm) {$z_3$};
            \else
              \ifnum\x=\NumDots
                \node at (\Angle:1.8cm) {$z_N$};
              \else
                \ifnum\x=\numexpr\NumDots-1\relax
                  \node at (\Angle:1.9cm) {$z_{N-1}$};
                \fi
              \fi
            \fi
          \fi
        \fi
      }
      \node[rotate=110] at (1.7,0.6) {...};
      \node[rotate=70] at (-1.7,0.6) {...};
    \end{tikzpicture}
    \caption{}
    \label{fig:subfiga}
  \end{subfigure}
  \hspace{10pt} 
  \begin{subfigure}[b]{0.4\textwidth}
    \centering
    \begin{tikzpicture}
      \draw (0,0) circle (1.5cm); 
      
      \newcommand{\NumDots}{17} 
      
     \foreach \x in {1,2,...,\NumDots} {
  \pgfmathsetmacro{\Angle}{90 - 360/\NumDots*(\x-1)}
  \ifnum\x=6
  \else
    \ifnum\x=13
    \else
      \fill (\Angle:1.5cm) circle (1.5pt);
      \ifnum\x=1
        \node at (\Angle:1.8cm) {$z_1$};
      \else
        \ifnum\x=2
          \node at (\Angle:1.8cm) {$z_2$};
        \else
          \ifnum\x=3
            \node at (\Angle:1.8cm) {$z_3$};
          \else
            \ifnum\x=\NumDots
              \node at (\Angle:1.8cm) {$z_N$};
            \else
              \ifnum\x=\numexpr\NumDots-1\relax
                \node at (\Angle:1.9cm) {$z_{N-1}$};
              \fi
            \fi
          \fi
        \fi
      \fi
    \fi
  \fi
}
      \node[rotate=110] at (1.7,0.6) {...};
      \node[rotate=70] at (-1.7,0.6) {...};
      \fill[red] (0.7,-0.2) circle (1.5pt);
      \node at (0.7+0.2,-0.2+0.25) {$u_1$};
       \fill[blue] (-1.9,-0.6) circle (1.5pt);
        \node at (-1.9-0.2,-0.6+0.25) {$v_1$};
    \end{tikzpicture}
    \caption{}
    \label{fig:subfigb}
  \end{subfigure}
  \caption{(a) In the large $N$ saddle point, the eigenvalues $z_i$ are distributed on the unit circle. (b) An instanton configuration where two eigenvalues are pulled off the unit circle.}
  \label{fig:mainfigure}
\end{figure}
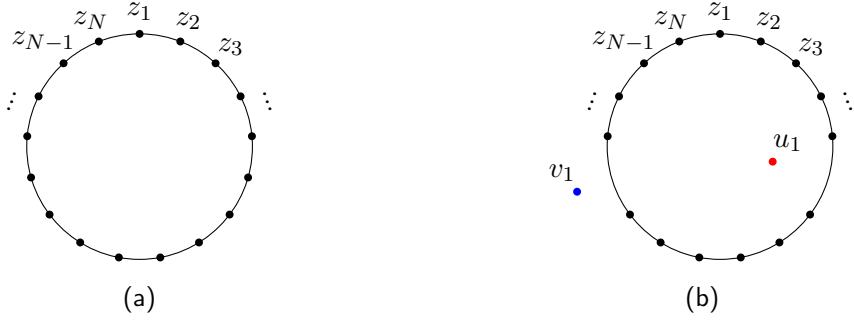

For now, we assume that the $t_k$'s are of order $N$ but small enough that we are in the phase where the Vandermonde repulsion dominates over the attraction due to the potential, and so the support of the eigenvalue distribution is the whole unit circle.
Let $z_i = e^{\i \theta_i}$, with $i = 1,\ldots, N$ denote the eigenvalues of $U$.
Explicitly, the eigenvalue density is given by 
\begin{align}
    \rho(\theta) = \frac{1}{2\pi} \left( 
    1 + \frac{1}{N} \sum_{k=1}^\infty \left( t_k^+ e^{\i k \theta} + t_k^- e^{-\i k \theta}\right)
    \right)\, ,
\end{align}
see Figure \ref{fig:subfiga} for an illustration.

A class of well-studied nonperturbative effects in matrix integrals are the so-called eigenvalue instanton effects \cite{Neuberger:1980qh, David:1990sk, Shenker:1990uf, Marino:2007te, Marino:2008ya}. 
The quantity controlling these is the effective potential $V_{\textrm{eff}}(z)$ felt by a single eigenvalue, which is a combination of the explicit potential in the matrix and the Vandermonde repulsion term.
To analyze these effects, it is useful to write the integral (\ref{zntildedef}) in the following form: 
\begin{align}
    \widetilde{Z}_N(\mathbf{t^+},\mathbf{t^-}) = 
    (-1)^\frac{N(N-1)}{2}\int \prod_{i=1}^N\frac{\d z_i}{2\pi \i z_i^N} \, \prod_{1 \leq i < j \leq N} (z_i - z_j)^2 
    \exp \left(- \sum_{k=1}^\infty V(z_k)
    \right)\, , \label{zntholo}
\end{align}
where $z_1,\ldots, z_N$ are the eigenvalues of $U$, and, for ease of later notation, we have defined
\begin{align}
    V(z) := - \sum_{k=1}^\infty \left( 
    \frac{t_k^+}{k} \sum_j z_j^k
    + \frac{t_k^-}{k} \sum_j z_j^{-k}
    \right)\, .
    \label{defv}
\end{align}
Let us also define
\begin{align}
    \widetilde{V}(z) := - \sum_{k=1}^\infty \left( 
    \frac{t_k^+}{k} \sum_j z_j^k
    - \frac{t_k^-}{k} \sum_j z_j^{-k}
    \right)\, ,
    \label{defvtilde}
\end{align}
a quantity that will also be useful later.

Since, in the present case, the support of the eigenvalue density divides the eigenvalue plane into two disconnected pieces, various quantities of interest in the matrix integral define two separate analytic functions of $z$, depending on whether we start outside or inside the unit circle in the complex-$z$ plane.
For instance, 
\begin{align}
    \langle \Tr \log (1 - z^{-1} U) \rangle_{(\mathbf{t^+},\mathbf{t^-})} &= - \sum_{j=1}^\infty 
    \frac{t_j^{-}}{j}\, z^{-j} \,, \quad \text{if } \vert z \vert > 1\, , \label{trace1}\\
    \langle \Tr \log (1 - z U^{-1}) \rangle_{(\mathbf{t^+},\mathbf{t^-})} &= - \sum_{j=1}^\infty 
    \frac{t_j^{+}}{j}\, z^{j} \,, \quad \text{if } \vert z \vert < 1\, , \label{trace2}
\end{align}
where the expectation values are taken using the integral (\ref{zntildedef}).
The way to derive these is to simply Taylor expand the function on the left hand side and then compute $\langle \Tr U^m \rangle_{(\mathbf{t^+},\mathbf{t^-})}$ by Taylor expanding the exponential in the integrand (\ref{zntildedef}) and using the following result from \cite{diaconis1994eigenvalues} to perform the resulting integral over $U(N)$:
\begin{align}
    \int \d U \, \prod_{j=1}^k 
    (\Tr U^j)^{a_j} (\Tr U^{-j})^{b_j} &= \prod_{j=1}^k j^{a_j} (a_j)!\, \delta_{a_j, b_j} 
    \quad\quad \text{if } N \geq \sum_{j=1}^k j \, a_j\, .
    \label{diacshah}
\end{align}
See, for example, \cite{Eniceicu:2023cxn} for more details.
The right hand sides of these equations can now be analytically continued outside their original domains of definition, to the entire complex plane.

Similarly, we can define a one-eigenvalue effective potential $V_\text{eff}^+$ by starting from outside the unit circle, and a one-eigenvalue effective potential $V_\text{eff}^-$ by starting from inside the unit circle.
The explicit expressions are \cite{Eniceicu:2023cxn}
\begin{align}
    V_\text{eff}^+(z) = - N \log(-z) &- \sum_{k=1}^\infty \left( 
    \frac{t_k^+}{k} z^k - \frac{t_k^-}{k} z^{-k} 
    \right) \,, \label{veffplus}\\
    V_\text{eff}^-(z) =  N \log(-z) &+ \sum_{k=1}^\infty \left( 
    \frac{t_k^+}{k} z^k - \frac{t_k^-}{k} z^{-k} 
    \right) \,. \label{veffminus}
\end{align}
In the 't Hooft limit, the couplings $t_k$ are of order $N$ and the extrema of the effective potential lead to contributions of order $e^{-cN}$ to the partition function.
Note that the relative sign between $t_k^+$ and $t_k^-$ is flipped on the right sides of (\ref{veffplus}) and (\ref{veffminus}) compared to the original action appearing in (\ref{zntildedef}).
This happens because of the minus signs on the right hand sides of (\ref{trace1}) and (\ref{trace2}).
Note also that the effective potentials are closely related to the generating function appearing in the definition of $\widetilde{K}$ in (\ref{ktildegenfn}).

The main idea is that $\widetilde{G}^{(m)}_N$ comes from a configuration where $m$ of the eigenvalues have tunneled to an extrema of $V_\text{eff}^+$ and $m$ of the eigenvalues have tunneled to an extrema of $V_\text{eff}^-$.\footnote{
The only nonzero contributions are when we have an equal number of instantons associated to $V_\text{eff}^+$ and $V_\text{eff}^-$ \cite{Eniceicu:2023cxn, Chakrabhavi:2024szk}.
}
Let us denote the eigenvalues that have tunneled to an extrema of $V_\text{eff}^+$ by $\{u_1,\ldots, u_m\}$ and 
the eigenvalues that have tunneled to an extrema of $V_\text{eff}^-$ by $\{v_1,\ldots, v_m\}$, see Figure \ref{fig:subfigb}.\footnote{
Note that even though we motivated the instanton configuration by thinking of $u_i$ and $v_i$ being at the extrema of the effective potentials, we are doing the full contour integral over these variables.
}
Denote the remaining $N-2m$ eigenvalues by $\{z_1, \ldots, z_{N-2m}\}$.
Following the procedure described in appendix C of \cite{Eniceicu:2023cxn}, we can write this contribution as 
\begin{align}
    &\widetilde{Z}_N^{(m|m)}(\mathbf{t^+},\mathbf{t^-})
    = \frac{(-1)^{\frac{1}{2}N(N-1)}}{N!}{N \choose m}{N-m \choose m} \,
    \int \prod_{i=1}^m \frac{\d u_i}{2\pi \i u_i^N}\frac{\d v_i}{2\pi \i v_i^N}
    \prod_{i=1}^{N-2m} \frac{\d z_i}{2\pi \i z_i^N} \times  \nonumber \\
    &\times \prod_{1\leq i<j\leq m}(u_i - u_j)^2(v_i - v_j)^2
    \prod_{i=1}^m\prod_{j=1}^m (u_i - v_j)^2
    \prod_{i=1}^m \prod_{j=1}^{N-2m} (u_i - z_j)^2(v_i - z_j)^2 
    \prod_{1\leq i<j\leq N-2m}(z_i - z_j)^2\times
    \nonumber \\
    & \times \exp \left(- \sum_{i=1}^m (V(u_i) + V(v_i)) - \sum_{i=1}^{N-2m} V(z_i) \right)\, .
    \label{ztildemmstep1}
\end{align}
The only thing that has been done in this step is to incorporate the correct combinatorial factors for pulling out the set of eigenvalues as described above, and to separate the Vandermonde factor into various pieces.
Next, we slightly rearrange the Vandermonde factors
\begin{align}
    &\widetilde{Z}_N^{(m|m)}(\mathbf{t^+},\mathbf{t^-})
    = \frac{(-1)^{\frac{1}{2}N(N-1)}}{(N-2m)!(m!)^2} \,
    \int \prod_{i=1}^m \frac{\d u_i}{2\pi \i u_i^{-N+4m}}\frac{\d v_i}{2\pi \i v_i^N}
    \prod_{i=1}^{N-2m} \frac{\d z_i}{2\pi \i z_i^{N-2m}} \times  \nonumber \\
    &\times \prod_{1\leq i<j\leq m}(u_i - u_j)^2(v_i - v_j)^2
    \prod_{i=1}^m\prod_{j=1}^m (u_i - v_j)^2
    \prod_{i=1}^m \prod_{j=1}^{N-2m} \left(1- \frac{z_j}{u_i} \right)^2 \left(1 - \frac{v_i}{z_j}\right)^2 
    \prod_{1\leq i<j\leq N-2m}(z_i - z_j)^2\times
    \nonumber \\
    & \times \exp \left(- \sum_{i=1}^m (V(u_i) + V(v_i)) - \sum_{i=1}^{N-2m} V(z_i) \right)\, .
    \label{ztildemmstep2}
\end{align}
Next, we just do the integral over the $z_i$'s, which amounts to computing a specific correlator in the matrix integral (\ref{zntholo}), apart from the fact that we have $N-2m$ eigenvalues instead of $N$.
We need the results (\ref{trace1}) and (\ref{trace2}), and also the following connected correlators \cite{Eniceicu:2023cxn}:
\begin{align}
    \langle \Tr \log(1-z_1^{-1}U)\Tr \log(1-z_2^{-1} U) \rangle_{(\mathbf{t^+},\mathbf{t^-})}
    &= 0 \quad\hspace{0.98in} \text{if } \vert z_1 \vert > 1 \text{ and } \vert z_2 \vert > 1\,, \\
    \langle \Tr \log(1-z_1^{-1}U)\Tr \log(1-z_2 U) \rangle_{(\mathbf{t^+},\mathbf{t^-})}
    &= -\log(1-z_2/z_1) \quad \text{if } \vert z_1 \vert > 1 \text{ and } \vert z_2 \vert < 1\,,\\
    \langle \Tr \log(1-z_1 U^{-1})\Tr \log(1-z_2 U^{-1}) \rangle_{(\mathbf{t^+},\mathbf{t^-})}
    &= 0 \quad\hspace{0.98in} \text{if } \vert z_1 \vert < 1 \text{ and } \vert z_2 \vert < 1\,,
\end{align}
with the expectation values taken in the integral (\ref{zntildedef}).
The result is
\begin{align}
    \widetilde{Z}_N^{(m|m)}(\mathbf{t^+},\mathbf{t^-})
    &= \widetilde{Z}_\infty(\mathbf{t^+},\mathbf{t^-})\, \frac{(-1)^m}{(m!)^2} \,
    \int \prod_{i=1}^m \frac{\d u_i}{2\pi \i u_i^{-N+4m}}\frac{\d v_i}{2\pi \i v_i^N} \times  \nonumber \\
    &\quad \times 
    \prod_{1\leq i<j\leq m}(u_i - u_j)^2(v_i - v_j)^2
    \prod_{i=1}^m\prod_{j=1}^m (u_i - v_j)^2
    \exp \left( - \sum_{i=1}^m \widetilde{V}(u_i) + \sum_{i=1}^m \widetilde{V}(v_i) \right) \times \nonumber \\
    &\quad \times \prod_{i=1}^m\prod_{j=1}^m \exp (- 4 \log(1 - v_j/u_i)) \nonumber\\
    &= \widetilde{Z}_\infty(\mathbf{t^+},\mathbf{t^-})\, \frac{(-1)^m}{(m!)^2} \,
    \int \prod_{i=1}^m \frac{\d u_i}{2\pi \i u_i^{-N}}\frac{\d v_i}{2\pi \i v_i^N} \times  \nonumber \\
    &\quad \times \frac{\prod_{1\leq i<j\leq m}(u_i - u_j)^2(v_i - v_j)^2}
    {\prod_{i=1}^m\prod_{j=1}^m (u_i - v_j)^{2}}
    \exp \left( - \sum_{i=1}^m \widetilde{V}(u_i) + \sum_{i=1}^m \widetilde{V}(v_i) \right)\, .\label{zmmt}
\end{align}
Note the appearance of $\widetilde{V}$ in the exponent above, which arises because of (\ref{trace1}) and (\ref{trace2}).
(Recall that $\widetilde{V}$ was defined in (\ref{defvtilde}). Compared to the potential $V$ (\ref{defv}), it has the opposite sign for the terms in the potential proportional to $t_k^-$.)
The presence of the factors of $u_i^{N}$ and $v_i^{-N}$ in the integrand of (\ref{zmmt}) is the reason why the effective potentials in (\ref{veffplus}) and (\ref{veffminus}) are taken to have a logarithmic term.
If one is interested in the matrix integrals with 't Hooft scaling of the couplings, one would proceed to find the extrema of the effective potential and proceed as in \cite{Eniceicu:2023cxn}.
However, our goal is different: We don't want to explicitly perform the $u,v$ integrals at this stage. 
We will keep the $u,v$ integrals as they are and will eventually arrive at the expression (\ref{danresult}).
Another important result from \cite{Eniceicu:2023cxn} is that the contributing saddle points for $u_i$ lie at $\vert z \vert < 1$, that is, they lie in the region beyond the original domain of definition of $V_\text{eff}^+$.
A similar comment applies for the $v_i$.
In this sense, we are discussing a contribution coming from $2m$ \emph{ghost instantons}. 
See \cite{Marino:2022rpz} for a detailed discussion of ghost instantons.

\paragraph{Doing the integrals over $\mathbf{t^+}$ and $\mathbf{t^-}$ variables.}
The main insight of this note into connecting the above calculation to the integral (\ref{zndef}) that computes the partition functions and indices of gauge theories is that, since the quantity $\widetilde{V}$ appearing in the exponent in (\ref{zmmt}) is linear in the variables $\mathbf{t^+}$ and $\mathbf{t^-}$, and $\widetilde{Z}_\infty(\mathbf{t^+},\mathbf{t^-})$ is the exponential of a term quadratic in these variables, the integral in the Hubbard-Stratonovich transformation (\ref{hstransformation}) can be explicitly performed.
In other words
\begin{align}
    &Z_\infty(\mathbf{g}) \,G^{(m)}_N(\mathbf{g})
    = \frac{(-1)^m}{(m!)^2} \,
    \int \prod_{i=1}^m \frac{\d u_i}{2\pi \i u_i^{-N}}\frac{\d v_i}{2\pi \i v_i^N} \prod_{1\leq i<j\leq m}(u_i - u_j)^2(v_i - v_j)^2
    \prod_{i=1}^m\prod_{j=1}^m (u_i - v_j)^{-2} \times  \nonumber\\
    &\times \int \prod_{k=1}^\infty \frac{\d t_k^+\d t_k^-}{2\pi k g_k} \exp \left( 
    - \frac{t_k^+ t_k^-}{k g_k}
    \right)
    \exp \left( 
    \frac{t_k^+ t_k^-}{k} \right)
    \exp \left( 
    - \frac{t_k^{+}}{k} (\sum_i u_i^k -  \sum_i v_i^k)
    + \frac{t_k^{-}}{k} (\sum_i u_i^{-k} -  \sum_i v_i^{-k})
    \right) \nonumber
\end{align}
Because of the comments relating to the discussion of ghost instantons above, the integration contour for each $u_i$ is a circle of radius smaller than one, and the integration contour for each $v_i$ is a circle of radius greater than one. 
The Gaussian integral over $\{t_k^+ ,t_k^-\}$ is easily performed, yielding
\begin{align}
    G^{(m)}_N(\mathbf{g})
    &= \frac{(-1)^m}{(m!)^2} \,
    \int \prod_{i=1}^m \frac{\d u_i}{2\pi \i u_i^{-N}}\frac{\d v_i}{2\pi \i v_i^N} \prod_{1\leq i<j\leq m}(u_i - u_j)^2(v_i - v_j)^2
    \prod_{i=1}^m\prod_{j=1}^m (u_i - v_j)^{-2} \times  \nonumber\\
    &\quad \times \exp \left( - \sum_{k=1}^\infty \frac{1}{k} \frac{g_k}{1-g_k} 
    (\sum_i u_i^k -  \sum_i v_i^k)
    (\sum_i u_i^{-k} -  \sum_i v_i^{-k})
    \right)\, .\label{gnresultfinal}
\end{align}
This precisely equals the expression (\ref{danresult}) for $G^{(m)}_N$ derived in \cite{Eniceicu:2023uvd}.
It can be interpreted as an integral over an $(m|m)$ supermatrix, with an insertion of the $N$-th power of the Berezinian.

\paragraph{Getting the giant graviton expansion without the Hubbard-Stratonovich transformation.}
Having understood the perspective above, we can now provide a more direct way to get the expansion (\ref{gnexpansion}), (\ref{danresult}) without the intermediate step of the Hubbard-Stratonovich transformation (\ref{hstransformation}), (\ref{zntildedef}).
We simply pull out two sets of $m$ eigenvalues $\{u_1, \ldots, u_m\}$ and $\{v_1, \ldots, v_m\}$ from the original double-trace integral (\ref{zndef}).
The integration measure and the Vandermonde factor still split up as the first and second line of (\ref{ztildemmstep2}).
However, letting $\tilde{U}$ denote the matrix with eigenvalues $\{z_1, \ldots, z_{N-2m}\}$, the double trace potential is now decomposed as follows
\begin{align}
    \Tr U^k \Tr U^{-k}
    &= \left(
    \sum_{i=1}^m u_i^k + \sum_{i=1}^m v_i^k 
    \right)
    \left(\sum_{i=1}^m u_i^{-k} + \sum_{i=1}^m v_i^{-k}\right) + \nonumber\\ 
    &\quad + \Tr \tilde{U}^k \left(\sum_{i=1}^m u_i^{-k} + \sum_{i=1}^m v_i^{-k}\right)
    + \Tr \tilde{U}^{-k} \left(\sum_{i=1}^m u_i^{k} + \sum_{i=1}^m v_i^{k}\right)
    + \Tr \tilde{U}^k \Tr \tilde{U}^{-k}
    \label{doubletraceuvzsplit}
\end{align}
The factors of $\det(1-u_i^{-1}\tilde{U})^2$ and $\det(1-v_i\tilde{U}^{-1})^2$ from the Vandermonde in the second line of (\ref{ztildemmstep2}) can be written as 
\begin{align*}
    \det(1-u_i^{-1}\tilde{U})^2 = \exp \left( 
    - 2 \sum_{k=1}^\infty \frac{u_i^{-k}}{k} \Tr \tilde{U}^k 
    \right)
\end{align*}
and then these factors can be combined with the corresponding terms in (\ref{doubletraceuvzsplit}).
The integral over $\widetilde{U}$ can now be done using the following expectation value in the integral (\ref{zndef}):
\begin{align}
    \left\langle\exp\left[\sum_k\frac{1}{k}\lambda_k^-\tr U^{k}+\frac{1}{k}\lambda_k^+\tr U^{-k}\right]\right\rangle_\mathbf{g}
    = \exp\left(\sum_k\frac{1}{k}\frac{\lambda_k^-\lambda_k^+}{1-g_k}\right)\, .
\end{align}
The way to derive this is again to expand the exponentials and use the result (\ref{diacshah}) from \cite{diaconis1994eigenvalues}.
So we get
\begin{equation}
\begin{aligned}
\frac{Z^{(m|m)}(\mathbf{g})}{Z_{\infty}(\mathbf{g})}&=\frac{(-1)^m}{(m!)^2}\int\prod_i\frac{\d u_i}{2\pi \i}\frac{\d v_i}{2\pi \i}\prod_i u_i^{N-4n}v_i^{-N}\times\prod_{i<j}(u_i-u_j)^2\prod_{i<j}(v_i-v_j)^2\prod_{i,j}(u_i-v_j)^2\\
&\times\exp\left[\sum_k\frac{g_k}{k}\left(\sum_i u_i^k+\sum_i v_i^k\right)\left(\sum_i u_i^{-k}+\sum_i v_i^{-k}\right)\right]\\
&
\times\exp\left[\sum_k\frac{1}{k}\frac{1}{1-g_k}\left((g_k-2)\sum_iu_i^{-k}+g_k\sum_iv_i^{-k}\right)\left(g_k\sum_iu_i^{k}+(g_k-2)\sum_iv_i^{k}\right)\right]\, .
\end{aligned}
\end{equation}
After some simple simplfications we recover the result (\ref{danresult}). 
Hence we have shown that $Z^{(m|m)}(\mathbf{g}) = G^{(m)}_N(\mathbf{g})$ without needing the intermediate Hubbard-Stratonovich transformation to the integral (\ref{zntildedef}).

Let us now consider two example integrals.

\paragraph{A toy model.} 
Let us consider the following toy integral with just one coupling $g_1$ turned on. 
This is a truncated version of the general integral (\ref{zndef}) and was studied, for example, in \cite{Copetti:2020dil}.
\begin{align}
    Z_N(g_1) := \int \d U \, \exp \left( 
    g_1 \Tr U \Tr U^{-1}
    \right)\, .
    \label{zngdef}
\end{align}
The Hubbard-Stratonovich transformation of this integral to the form (\ref{zntildedef}) is the original Gross-Witten-Wadia model \cite{Gross:1980he, Wadia:1980cp, Wadia:2012fr}.
Applying the expansion (\ref{gnexpansion}) using the explicit expression (\ref{gnresultfinal}) for the $m=1$ contribution, we find
\begin{align}
    G^{(1)}_N(g_1) = -\int \frac{\d u}{2\pi \i} \frac{\d v}{2\pi \i} \frac{u^N}{v^N} \,\frac{1}{(u-v)^2} \, \exp \left( - \frac{g_1}{1-g_1} (2 - u/v - v/u) \right)\, .
    \label{gn1toy}
\end{align}
We will first do the $u$ integral.
The contour for $u$ is inside the unit circle and that for $v$ is outside.
We would like to deform the $u$ contour towards the origin.
and so, because of the presence of the high positive power $u^N$, we need a term proportional to $u^{-N-1}$ from expanding the exponential.
The first such term is
\begin{align}
    -\int \frac{\d v}{2\pi \i v^N}  \frac{\d u \, u^N}{2\pi \i}\,  \frac{1}{(u-v)^2}\,
    \frac{1}{(N+1)!}\left( - \frac{g_1}{1-g_1}\right)^{N+1}
    \left(- \frac{v}{u} \right)^{N+1}
    &= -\frac{g_1^{N+1}}{(N+1)!} \int \frac{\d v}{2\pi \i \, v} + O(g_1^{N+2}) \, . \nonumber
\end{align}
Using this on the right hand side of (\ref{gn1toy}), we get
\begin{align}
    G^{(1)}_N(g_1) &= -\frac{g_1^{N+1}}{(N+1)!} + O(g_1^{N+2}) \, .
    \label{G1g1}
\end{align}
This is the known $m=1$ ``giant graviton'' contribution to this toy integral \cite{Murthy:2022ien, Eniceicu:2023cxn}. 
A striking feature of this result is the factorial dependence on $N$, i.e., the leading behavior is $\exp(- N \log N)$. 
This does not happen for more general integrals of the form (\ref{zndef}) where infinitely many couplings are turned on. 
In fact, it is possible to reproduce the result (\ref{G1g1}), with the  $\frac{1}{(N+1)!}$ factorial replaced by its Stirling approximation, by using the perturbative plus two-(ghost)instanton approximation to the partition function of the Gross-Witten-Wadia integral in the ungapped phase \cite{Marino:2008ya, Ahmed:2017lhl, Eniceicu:2023cxn}, and performing the integral over $t_1^+, t_1^-$ in (\ref{hstransformation}).
Thus, in this case, we can really think of the ``giant graviton'' corrections as arising from eigenvalue \emph{saddle-points}.
However, in the general case of the integral (\ref{zndef}) where infinitely many couplings are turned on, the $u$ and $v$ integrals are not amenable to the saddle-point approximation.

\paragraph{The $1/2$-BPS index in $\mathcal{N}=4$ SYM.}
The $1/2$-BPS index of $\mathcal{N}=4$ SYM \cite{Kinney:2005ej} is a much studied example in the context of the giant graviton expansion \cite{Gaiotto:2021xce, Lee:2023iil, Murthy:2022ien, Eniceicu:2023uvd}.
In this case the couplings are given by $g_k = q^k$.
In the expression for the $m=1$ giant graviton (\ref{gnresultfinal}), we can expand $\frac{g_k}{1-g_k} = \frac{q^k}{1-q^k} = \sum_{n=1}^\infty q^{kn}$ and rearrange the sum in the exponential of (\ref{gnresultfinal}) to simplify
\begin{align}
    G^{(1)}_N(q) = - \int \frac{\d u}{2\pi \i} \frac{\d v}{2\pi \i} \frac{u^N}{v^N} \frac{1}{(u-v)^2} \prod_{n=1}^\infty \frac{(1-q^n)^2}{(1-\frac{u q^n}{v})(1-\frac{v q^n}{u})}\, .
\end{align}
Assuming, for simplicity, that $q \ll 1$, we can first do the $u$ integral by deforming the contour towards the origin and pick up the pole at $u = v q$ coming from the second term in the denominator with $n=1$. 
This gives
\begin{align}
    G^{(1)}_N(q) &= - \int \frac{\d v}{2\pi i} \, \frac{(vq)^N}{v^N} \frac{1}{(vq-v)^2}\, v q\, 
    \prod_{n=1}^\infty \frac{(1-q^n)^2}{(1-q^{n+1})^2} \prod_{n=2}^\infty
    \frac{1}{1-q^{n-1}} + O(q^{N+2}) \nonumber \\
    &= - q^{N+1} \int \frac{\d v}{2\pi \i v} + O(q^{N+2})
    = - q^{N+1} + O(q^{N+2})\, .
\end{align}
This is indeed the known result \cite{Gaiotto:2021xce, Lee:2023iil, Murthy:2022ien, Eniceicu:2023uvd}.
The coefficient $-1$ in front of $q^{N+1}$ represents the absence of a single giant graviton with R-charge $N+1$, the first forbidden giant in the theory with finite $N$ \cite{McGreevy:2000cw}.
In other words, it signifies the appearance of the first trace relation.

\paragraph{Acknowledgments.}
We would like to thank Dan Stefan Eniceicu, Ji Hoon Lee, Chitraang Murdia, Douglas Stanford and especially Steve Shenker for useful discussions. 
Y.C. acknowledges support from DOE grant DE-SC0021085. 
H.T. would like to thank professor Xiao-Liang Qi's support by National Science Foundation under grant No.2111998 and the Simons Foundation. 

\small
\bibliographystyle{apsrev4-1long}
\bibliography{main}
\end{document}